\documentclass[a4j,12pt]{article}%
\usepackage{amsmath,amsthm,amssymb}
\usepackage{epic, eepic}
\usepackage{graphicx}%
\usepackage{amsmath}%
\usepackage{amsfonts}%
\usepackage{amssymb}
\providecommand{\U}[1]{\protect\rule{.1in}{.1in}}
\newtheorem{theo}{Theorem}[section]
\newtheorem{theor}{Theorem}[section]

\newtheorem{defi}{Definition}[section]
\newtheorem{rema}{Remark}[section]

\newtheorem{axio}{Axiom}
\newtheorem{axiom}{Axiom}

\begin{document}

\title{\vspace{-15mm} Derivation of Born Rule from \\Algebraic and Statistical Axioms  \vspace{-3mm}}
\author{Izumi Ojima\thanks{ojima@kurims.kyoto-u.ac.jp} \; and Kazuya Okamura\thanks{kazuqi@kurims.kyoto-u.ac.jp}\\
RIMS, Kyoto University\\
Hayato Saigo\thanks{h\_saigoh@nagahama-i-bio.ac.jp}\\
Nagahama Institute of Bio-Science and Technology}
\date{}
\maketitle
\vspace{-12mm}
\begin{center}
\textit{On the occasion of 300th Anniversary of Jacob Bernoulli's Ars Conjectandi}
\end{center}

\begin{abstract}
In the present paper we propose a new system of algebraic and statistical 
axioms as working hypotheses, from which Born rule can be seen to emerge. In this process 
the concept of 
sectors defined as quasi-equivalence classes of factor states plays a crucial role.  
\end{abstract}

\section{Introduction}

Entities in atomic scale have intrinsic indeterminancy and ``statistical'' description is more or less inevitable in quantum theory. 
So it is natural to think that the use of statistics as an ``Ars Conjectandi (art of conjecturing)''\cite{BeAC} 
is essential for physics.
However, one cannot obtain any fruitful viewpoints for quantum theory by just reading the textbooks on statistics. 
For instance, it seems almost impossible at first sight to relate mathematical statistics and 
``statistical'' aspects of quantum theory such as the Born rule \cite{Born}. Does it mean statistics is useless for fundamental study for physics? In the present paper we will show that the answer is NO: Statistical description naturally emerges
out of the algebraic framework for quantum theory.  

Statistics is not an isolated science from other sciences. 
As C.R.Rao \cite{RaST} says, statistics ``cannot be used in a routine way; the user
must acquire the necessary expertise to choose the right technique in
a given situation and make modifications, if necessary''.
The appropriate question is that how we can formulte statistics for Physics, and dually, Physics as Statistical Science. 
Why, and in which point, do statistical descriptions become crucial for quantum theory? 

Here let us fix the meaning of statistical to be based on the generic concept of random variables, i.e., variables 
whose fluctuation is subject to a situation. 
Kolmogorov \cite{KoGW} formulated the notion of random variables 
as functions on some sample spaces. Intuitively speaking, the set of all possible choices is determined ``in advance''. However, the development of quantum theory have unveiled that 
the Kolmogorovian formulation of random variables is too narrow to cover the quantities of atomic entities whose fluctuation is subject to a physical situation. 
In quantum theory the quantities with which we are concerned are often noncommutative and there exist no sample space
 on which all the quantities are defined as functions. 
Sample spaces for physical quantities should be considered as objects emerging out of each situations and determined \textit{a posteriori}. 

In the present paper we propose a new axiomatic system for quantum theory from the algebraic and statistical viewpoint. 
Here the term ``axiom'' should be interpreted as ``working hypothesis.'' The authors believe that this is similar to the 
original idea of Hilbert's\cite{HiAD}. The essence of this original meaning of axioms was nicely shared by the pioneers of axiomatic quantum field theory, 
whereas its common understanding has been deformed in such a wrong way as worshiping a kind of ``sacred canons''.

We adopt the following three axioms:
\begin{axiom}
All the statistical aspects of a physical system are registered in a C*-probablity space.
\end{axiom}

\begin{axiom}[Sector]
For a state $\omega \in E_{\mathcal{A}}$ and a Borel
set $\Delta \subset E_{\mathcal{A}}$, $d\omega (\Delta)$
gives the probability that a sector belongs to $\Delta$
under the situation described by $\omega$. When available
observables are restricted, the coarse-grained probability is
given by $d^{\mathcal{B}} \omega (\Delta)$ for some
subalgebra $\mathcal{B}$ of $\mathcal{Z}_\omega (\mathcal{A})$.
\end{axiom}

\begin{axiom}
The observable algebra for the composite system of an object system and a measuring apparatus is given by 
$\mathcal{A}\otimes C_0(M)  
(\subset \pi_\omega(\mathcal{A})^{\prime\prime}\otimes L^\infty(M))$, $M=\widehat{\mathcal{U}_A}$ or $\mathbb{R}$. 
A measurement process is described 
by a pair $\mathfrak{m}=(\alpha_{\mathfrak{m}},\psi_\mathfrak{m})$ of $\alpha_{\mathfrak{m}}\in Aut(\pi_\omega(\mathcal{A})^{\prime\prime}\otimes L^\infty(M))$ and of a state $\psi_\mathfrak{m}$ of the measuring apparatus.
The state of the composite system ``after measurement'' is given by
\[
\varphi_\mathfrak{m}(X):=
(\tilde{\varphi}\otimes \psi_\mathfrak{m})(\alpha_\mathfrak{m}((\pi_\omega\otimes id)(X))).
\]
\end{axiom}

From our axiomatic system the Born rule below can be seen to emerge in parallel with the above picture\footnote{The original form of this formulation was reported in \cite{OOS12, Ok12}.
The idea of this investigation can be traced back to the efficient use of sector 
in ``Large Deviation Strategy" \cite{OO12a,OO12b}.}:

\begin{theo}[Born rule]\label{Born}
Assume that the state of the object system is $\varphi=\tilde{\varphi}\circ\pi_\omega$.                                                                                                                                                         Then the probabilty $\mathrm{Pr}\{A\in\Delta\Vert\varphi\}$ with which 
the value of $A\in\pi_\omega(\mathcal{X})^{\prime\prime}$ is 
in $\Delta\in\mathcal{B}(\mathbb{R})$ is given by 
\begin{equation*}
\tilde{\varphi}(E^A(\Delta))=\langle \xi_{\tilde{\varphi}},E^A(\Delta)\xi_{\tilde{\varphi}} \rangle.
\end{equation*}
\end{theo}

The notions and notations above are given in the following sections. In section 2 the generic concepts of random variable and a 
situation are formulated as elements of a (noncommutative) algebra and a linear functional on it, respectively (Axiom 1). 
The concept of sectors, which is crucial in the derivation of Born rule, is defined in Section 3 as ``quasi-equivalence class'' of factor states. 
Sectors can be understood as labels for macroscopic classification of microscopic structure and 
be considered as a generalized notion of ``phase.'' Any state can be decomposed 
into sectors weighted by a natural measure called ``central measure'', whose existence is ensured by 
Tomita decomposition theorem \cite{BR1}. On this basis we propose the second axiom that the central measure 
determines the probability for a pure phase to emerge 
in actual space(-time) out of 
the mixed phase. The latter is to be understood as probabilitistic mixture in 
the probability space consisting of sectors as ``elementary events''. 
For a concrete formulation, we introduce a new concept ``instrument functional'' which is a generalization of several notions in quantum measurement theory such as instruments \cite{DL, Oz84, Oz97}.
In section 4 we propose the third axiom for a measurement process in terms of an automorphism on the composite system of the 
observed system and the apparatus, equipped with a state of the apparatus ``before measurement''. We derive a generalization of Born rule from these axioms in Section 5 
and the usual Born rule in terms of ``spectral equivalence'' (originally formulated by Ozawa \cite{Oz05,Oz06} under the name of ``perfect correlation'' and generalized by KO in \cite{OjOk13}) in Section 6.      

\section{Quantum Probability}
In quantum theory, observables can be considered as random variables which are not necessarily commutative. 
One of the most natural and general formulation of the systems of such quantities is based on the notion of *-algebras.

\begin{defi}
Let $\mathcal{A}$ be an algebra. A map $^{\ast}:\mathcal{A}\rightarrow \mathcal{A}$ is called an involution if the following equalities
\[(A^{\ast})^{\ast}=A, \:\: (A+B)^{\ast}=A^{\ast}+B^{\ast}, \:\: (\lambda A)^{\ast}=\bar{\lambda}A^{\ast}, \:\: (AB)^{\ast}=B^{\ast}A^{\ast}
\]
hold for any $a\in\mathcal{A}$ and $\lambda\in \mathbb{C}$. 
An algebra equipped with an involution is called a *-algebra. 
A map between two (unital) *-algebras is said to be a *-morphism if it is a homomorphism between algebras and preserves the involution.
\end{defi}

We assume that algebras have units, otherwise mentioned. It is known that units can be added if necessary.
Then the concept of states as expectations fuctionals is defined as follows. 
\begin{defi}
Let $\mathcal{A}$ be a unital *-algebra. A linear functional $\omega : \mathcal{A}\rightarrow \mathbb{C}$ is called a state on $\mathcal{A}$ if
\[
\omega(A^{\ast}A)\geq 0
\]
holds for any $A\in \mathcal{A}$ and 
\[
\omega(1)=1.
\]
\end{defi}


\begin{defi}
A pair $(\mathcal{A},\omega)$ of a *-algebra $\mathcal{A}$ and a state $\omega$ on $\mathcal{A}$ is called an algebraic probability space. 
\end{defi}

Although this quite simple concept of ``probability space'' is already useful to deal 
with many interesting aspects in mathematics and physics \cite{HoOb}, 
we will focus on a subclass of algebraic probability spaces to discuss the topological aspects such as approximation.


\begin{defi}
A *-algebra $\mathcal{A}$ equipped with a norm $\|\cdot\|$ is called a C*-algebra if it is complete with respect to $\|\cdot\|$ and 
\[
\|A^{\ast}A\|=\|A\|^2
\]
holds for any $A\in\mathcal{A}$. An algebraic probability space $(\mathcal{A},\omega)$ is called a C*-probability space if $\mathcal{A}$ is a unital C*-algebra.
\end{defi}


\begin{axio}
All the statistical aspects of a physical system are registered in a C*-probablity space.
\end{axio}

When a C*-algebra algebra is commutative, the assosiated C*-probability space can be represented as a usual probability space (Riesz-Markov-Kakutani theorem \cite{HoOb}). 
In other words, 
for the commutative case a 
C*-probability space has the same information as a (measure theoretic) probability space. Apart from this aspect, the essential ingredient of the ``information'' of probability space should be representable by 
the concept of ``events.'' To recover such a fundamental concept in the noncommutative algebraic context, 
we need projections as ``Yes-No questions,'' which, in commutative case, corresponds to indicator functions. They are provided by means of 
representations of C*-algebras on Hilbert spaces constructed from C*-probability spaces.

Let $\mathbf{B}(\mathcal{H})$ be the *-algebra of all bounded operators on a Hilbert 
space $\mathcal{H}$, where the involution is Hermitian conjugate and the norm is given by the operator norm. 

A typical example of C*-probablity space is $(\mathbf{B}(\mathcal{H}), \langle\Omega, (\cdot)\Omega\rangle)$, where $\Omega$ is a unit vector in a Hilbert space $\mathcal{H}$.
On the other hand, every C*-probability space can be formulated in a similar manner by ``GNS representation'' as follows. 

\begin{defi}
Let $\mathcal{A}$ be a unital *-algebra and $\mathcal{H}$ a Hilbert space.  
A *-morphism $\pi : \mathcal{A}\rightarrow \mathbf{B}(\mathcal{H})$ is called a *-representation of 
$\mathcal{A}$ on $\mathcal{H}$.
\end{defi}


\begin{theo}[Gelfand-Naimark-Segal representation theorem]

Let $(\mathcal{A},\omega)$ be a C*-probability space. There exist a *-representation $\pi_{\omega}$ of  $\mathcal{A}$ on a Hilbert space $\mathcal{H}_{\omega}$ and a 
vector $\Omega_{\omega}\in \mathcal{H}_{\omega}$ such that 
\[
\omega(A)=\langle \Omega_{\omega}, \pi_{\omega}(A) \Omega_{\omega} \rangle
\]
and $\pi_{\omega}(\mathcal{A})\Omega_{\omega}:=\{\pi(A)\Omega_{\omega}| A \in \mathcal{A}\}$ is dense in $\mathcal{H}_{\omega}$.

\end{theo}

The *-representation $\pi_{\omega}:\mathcal{A}\rightarrow \mathbf{B}(\mathcal{H}_\omega)$ 
equipped with $\Omega_{\omega}\in \mathcal{H}_{\omega}$ stated in the theorem above, 
or equivallently, the triple $(\pi_{\omega}, \mathcal{H}_{\omega}, \Omega_{\omega} )$,  
is called GNS representation of $(\mathcal{A},\omega)$. A GNS representation is ``unique up to unitaries''.






In contrast to the previous comparison in the commutative case, 
general C*-probability spaces can be viewed as ``noncommutative measure spaces'' represented in Hilbert spaces. Indeed, a von Neumann algebra defined below as a
$\sigma$-weakly closed (or equivalently, weakly closed) *-subalgebra of $\mathbf{B}(\mathcal{H})$ is known to 
provide a generalized measure theoretic structure in terms of projections.
   
\begin{defi}
The $\sigma$-weak topology on $\mathbf{B}(\mathcal{H})$ is 
the weakest topolgy which makes all the linear functionals of the form $\Sigma \langle y_n, (\cdot ) x_n\rangle$ $(\Sigma\|x_n\|^2, \Sigma\|y_n\|^2< \infty)$.
\end{defi}

\begin{defi}
A unital *-subalgebra of $\mathbf{B}(\mathcal{H})$ is called a von Neumann algebra over $\mathcal{H}$ if 
it is closed with respect to the $\sigma$-weak topology. 
\end{defi}

The class of states inherent in von Neumann algebra is normal ones defined below.

\begin{defi}
A state $\omega$ on a von Neumann algebra $\mathcal{A}$ is said to be normal if it is continuous with respect to $\sigma$-weak topology.
\end{defi}



A von Neumann algebra can be characterized algebraically as follows.
\begin{defi}
Let $S$ be a subset of $\mathbf{B}(\mathcal{H})$. A subset $S'$ of $\mathbf{B}(\mathcal{H})$ defined as 
\[
S'=\{A\in \mathbf{B}(\mathcal{H}) |\: \forall X \in S \:\:\:AX=XA \}
\]
is called the commutant of $S$.
\end{defi}

\begin{theo}(von Neumann's double commutant theorem)
A unital *-subalgebra $\mathcal{A}$ of $\mathbf{B}(\mathcal{H})$ is a von Neumann algebra if and only if $\mathcal{A}''=\mathcal{A}$. 
\end{theo}



\section{The Concept of Sector}


Let $\mathcal{A}$ be a C*-algebra. We denote the set of all states on $\mathcal{A}$ by $E_{\mathcal{A}}$.
A state $\omega\in E_{\mathcal{A}}$ is called a factor state when the center $\mathcal{Z}_{\omega}(\mathcal{A})$ of the 
von Neumann algebra 
$\pi_{\omega}(\mathcal{A})^{\prime\prime}$ is trivial, i.e.,  
$\mathcal{Z}_{\omega}(\mathcal{A}):=\pi_{\omega}(\mathcal{A})^{\prime
\prime}\cap\pi_{\omega}(\mathcal{A})^{\prime}\cong \mathbb{C}$. 
The set of all 
factor states on $\mathcal{A}$ is denoted by $F_{\mathcal{A}}$.

Let $\pi:\mathcal{A}\rightarrow B(\mathcal{H})$ be a representation of $\mathcal{A}$. 
A state $\omega\in E_{\mathcal{A}}$ is called $\pi$-normal if there exists a normal state $\rho$ on 
$\pi(\mathcal{A})^{\prime\prime}$ and
\begin{equation}
\omega(A)=\rho(\pi(A))
\end{equation}
holds for any $A\in\mathcal{A}$. We denote such $\rho$ as $\tilde{\omega}$.

\begin{defi}\quad 
\begin{enumerate}
 \item Two representations $\pi_{1}$ and $\pi_{2}$ are said to be quasi-equivalent and denoted as $\pi
_{1}\approx\pi_{2}$ when any $\pi_{1}$-normal state is $\pi_{2}$-normal and vice versa. Two states are said to be quasi-equivalent if the GNS representations are quasi-equivalent.
 \item Two representations $\pi_{1}$ and $\pi_{2}$ are said to be disjoint if there is no nontrivial intertwiner $T$, that is,
  \[T\pi_{1}(X)=\pi_{2}(X)T\Rightarrow T=0\;(X\in\mathcal{A}).\]
  Two states are said to be disjoint if the GNS representations are disjoint.
\end{enumerate}
\end{defi}

Two factor states are disjoint if they are not quasi-equivalent \cite{BR1, Emch}. 

\begin{defi}
[\cite{Oj03}] We call a quasi-equivalence class of factor states on $C^{\ast}$-algebra $\mathcal{A}$ as a sector for $~\mathcal{A}$.
\end{defi}

As disjointness of states corresponds to macroscopic distinguishability, 
a sector plays a role of a label for  macroscopic classification and 
should be considered as a generalized notion of ``phase''.

The following is a collorary of the fundamental theorem on state decomposition \cite[Theorem 4.1.25]{BR1}.

\begin{theo}
[Collorary of Tomita Decomposition Theorem] Let $\mathcal{A}$ be a $C^{\ast}$-algebra and $\omega$ a state on $\mathcal{A}$. There is a 
one-to-one correspondence between the two sets below:

$(1)$ $\{$Borel measures $\mu$ such that $ \int_{\Delta}\rho\;d\mu(\rho)$  and $ \int_{E_{\mathcal{A}%
}\backslash \Delta}\rho\;d\mu(\rho)$ are disjoint for any Borel set $\Delta\subset
E_{\mathcal{A}}$$\}$; 

$(2)$ $\{$commutative von
Neumann algebras $\mathcal{B}\subset \mathcal{Z}_{\omega}(\mathcal{A})  
\}.$ 

The above $\mathcal{B}$ is *-isomorphic to the image of the map $\kappa_{\mu}: L^{\infty}(\mu):=L^{\infty}(E_\mathcal{A},\mu)\ni f\mapsto\kappa_{\mu}(f)\in\mathcal{Z}_{\omega}(\mathcal{A})$
defined by 
\[\langle\Omega_{\omega},\kappa_{\mu}(f) \pi_{\omega}(A)
\Omega_{\omega}\rangle= \displaystyle{\int} d\mu(\rho) f(\rho)\widehat{A}%
(\rho).\]
\end{theo}


\begin{defi}
The measures characterized in the theorem above are called subcentral measures for $\omega$. 
The subcentral measure for $\omega$ corresponding to $\mathcal{B}$ is denoted by $d^\mathcal{B}\omega$. The subcentral measure corresponding to $\mathcal{B}=\mathcal{Z}_{\omega}(\mathcal{A})$ itself is called the 
central measure for $\omega$ and also denoted as $d\omega$.

\end{defi}



Sectors belonging to the support of the central measure for $\omega$ are distingushed by 
$Spec(\mathcal{Z}_{\omega}(\mathcal{A}))$, the spectrum of $\mathcal{Z}_{\omega}(\mathcal{A})$, which are macroscopically observable in the form of order parameters. In other words, 
they are labeled by the spectrum of observables 
in the center.

Since any state can be decomposed into sectors by the theorem above, we propose the following axiom:

\begin{axio}[Sector]


For a state $\omega \in E_{\mathcal{A}}$ and a Borel
set $\Delta \subset E_{\mathcal{A}}$, $d\omega (\Delta)$
gives the probability that a sector belongs to $\Delta$
under the situation described by $\omega$. When available
observables are restricted, the coarse-grained probability is
given by $d^{\mathcal{B}} \omega (\Delta)$ for some
subalgebra $\mathcal{B}$ of $\mathcal{Z}_\omega (\mathcal{A})$.
\end{axio}


This means that the central measure corresponding to a state determines the probability for a pure phase to emerge 
in actual space(-time) out of 
the mixed phase, the latter of which is to be understood as probabilitistic mixture in 
the probability space consisting of sectors as ``elementary events''. 
Each sector is directly associated with the choice of a state just after measurement as we will see below. 
This means that classical probability emerges out of quantum probability in 
the composite system of the object system and the measuring apparatus.

For the concrete formulation we introduce a new concept, 
instrument functional: 

\begin{defi}[Instrument functional]

A map $\mathcal{I}((\cdot) ;d^\mathcal{B}\omega)(\cdot)$ defined as
\[
\mathcal{I}(f;d^\mathcal{B}\omega)(X)=\int d^\mathcal{B}\omega(\rho)\;f(\rho)\;\rho(X),
\hspace{4mm}f\in L^\infty(d^\mathcal{B}\omega), X\in\mathcal{A},
\]
is called the instrument functional for $d^\mathcal{B}\omega$.

\end{defi}

By definition it holds that
\[
\mathcal{I}(f;d^\mathcal{B}\omega)(X)=
\langle \Omega_\omega, \kappa_{d^\mathcal{B}\omega}(f)\pi_\omega(X)\Omega_\omega \rangle.
\]


The concept of instrument functional is a generalization
of several notions in quantum measurement theory such as instruments \cite{DL, Oz84, Oz97}, 
which describes simultaneously two aspects in quantum measurement of distinguishing by 
$f\in L^{\infty}(d^\mathcal{B}\omega)$ and outputs of $X \in \mathcal{A}$ in each sector.

Especially, in the case that $\mathcal{B}=\mathcal{Z}_{\omega}(\mathcal{A})$, $X=1$ 
and $f=\chi_{\Delta}$ (indicator function of $\Delta$), 
it gives $d^\mathcal{B}\omega(\Delta) $ and by the axiom above it is equal to the probability that a sector is in $\Delta$.


\section{Measurement Process}

We shall formulate measurement processes in terms of composite systems. 
A measurement interaction is defined as an automorphism on an algebra of a 
composite system of an object system and an apparatus, equipped with a reference state on the object system ``before measurement''. 
Then we define a measurement process as a pair of an automorphism and a state on the apparatus.

Let $\mathcal{A}$ be an observable of an object system and $\omega$ be a reference state.
$(\pi_\omega,\mathcal{H}_\omega,\Omega_\omega)$ denotes the GNS representation of $\mathcal{A}$ with respect to $\omega$.
For simplicity, we consider the case of measuring $A\in\pi_\omega(\mathcal{A})^{\prime\prime}$. 

The spectrum of an observable $A$ of the appratus can be embedded into $\widehat{\mathcal{U}_A}$ (or $\mathbb{R})$\footnote{Because of its universality, $\mathbb{R}$ can be chosen as the spectrum for convenience' sake. }, where $\widehat{\mathcal{U}_A}$ denotes the dual group 
of $\mathcal{U}_A=\{u\in\pi_\omega(\mathcal{A})^{\prime\prime}\:|\:u=e^{itA}, t\in\mathbb{R}\}$. 
$C_0(\widehat{\mathcal{U}_A})$ (or $C_0(\mathbb{R})$) can be chosen as the algebra for the 
apparatus, where $C_0(M)$ denotes the algebra of 
continuous function on a locally compact (Hausdorff) space $M$ vanishing at infinity.
We adopt $\mathcal{A}\otimes C_0(M) ( \subset \pi_\omega(\mathcal{A})^{\prime\prime}\otimes 
L^\infty(M))$\footnote{$L^\infty(\widehat{\mathcal{U}_A})=L^\infty(\widehat{\mathcal{U}_A}, d\gamma)$, where $d\gamma$ denotes the Haar measure of $\widehat{\mathcal{U}_A}$.}, $M=\widehat{\mathcal{U}_A}$ or $\mathbb{R}$, as an 
observable algebra for the composite system of the object system and the measuring apparatus.

Then we assume that a measurement interaction is described by 
$\alpha\in Aut(\pi_\omega(\mathcal{A})^{\prime\prime}\otimes L^\infty(M))$, $M=\widehat{\mathcal{U}_A}$ or $\mathbb{R}$.
For any $X\in \mathcal{A}\otimes C_0(\widehat{\mathcal{U}_A})$, $\alpha((\pi_\omega\otimes id)(X))$ gives 
``$X$ after measurement of $A$''.

To reinterpret this situation with $X$ unchanged, it is more convenient to move to the 
``dual'' picture in terms of states, instead of focusing on observables. 

Summing up our discussions, we propose  




\begin{axio}
The observable algebra for the composite system of an object system and a measuring apparatus is given by 
$\mathcal{A}\otimes C_0(M)  
(\subset \pi_\omega(\mathcal{A})^{\prime\prime}\otimes L^\infty(M))$, $M=\widehat{\mathcal{U}_A}$ or $\mathbb{R}$. 
A measurement process is described 
by a pair $\mathfrak{m}=(\alpha_{\mathfrak{m}},\psi_\mathfrak{m})$ of $\alpha_{\mathfrak{m}}\in Aut(\pi_\omega(\mathcal{A})^{\prime\prime}\otimes L^\infty(M))$ and of a state $\psi_\mathfrak{m}$ of the measuring apparatus.
The state of the composite system ``after measurement'' is given by
\[
\varphi_\mathfrak{m}(X):=
(\tilde{\varphi}\otimes \psi_\mathfrak{m})(\alpha_\mathfrak{m}((\pi_\omega\otimes id)(X))).
\]
\end{axio}

In the case of $M=\mathbb{R}$ where a measuring pointer runs,
a measurement process of the ideal measurement of an observable $A$ is nothing but 
the pair of the measuring interaction $Ad\;e^{i\gamma(A\otimes P)}$ of von Neumann type \cite{vNQM,Oz88},
and of the delta measure $\delta_0$ as a state on $L^\infty(\mathbb{R})$.
On the other hand, in the case of $M=\widehat{\mathcal{U}_A}$, 
a measurement process of the ideal measurement of an observable $A$ is specified by
the pair $(\tau_U(V), \delta_\iota)$, where $\delta_\iota$ is the Dirac measure on a unit element
$\iota$ of $\widehat{\mathcal{U}_A}$, and $\tau_U(V)$ is an automorphism of
$\pi_\omega(\mathcal{X})^{\prime\prime}\otimes L^\infty(\widehat{\mathcal{U}_A})$
defined as follows: According to SNAG (Stone-Naimark-Ambrose-Godement) theorem \cite{StWi}, 
there exists a projection-valued measure (PVM) $E_U$ of $U(\mathcal{U}_A)^{\prime\prime}$
with the value space $(\widehat{\mathcal{U}_A},\mathcal{B}(\widehat{\mathcal{U}_A}))$
such that
\begin{equation*}
U_u=\int_{\widehat{\mathcal{U}_A}} \overline{\gamma(u)}\;dE_{U}(\gamma),\hspace{5mm}u\in \mathcal{U}_A,
\end{equation*}
for a unitary representation $(\mathcal{H}_\omega,U_u)$ of a locally compact abelian group $\widehat{\mathcal{U}_A}$
defined by $U_u \xi=u\xi$ for every $\xi\in\mathcal{H}_\omega$ ($u\in \mathcal{U}_A$).
Using $E_{U}$, we define a unitary operator $E_{U}(V)$
on $\mathcal{H}_\omega\otimes L^2(\widehat{\mathcal{U}_A})$ by
\begin{equation*}
E_{U}(V)=\int_{\widehat{\mathcal{U}_A}}dE_{U}(\gamma)\otimes\hat{\lambda}_{\gamma},
\end{equation*}
where $\hat{\lambda}_{\gamma}$ is the regular representation of $\widehat{\mathcal{U}_A}$
on $L^2(\widehat{\mathcal{U}_A})$ defined by 
$(\hat{\lambda}_{\gamma}f)(\gamma^\prime)=f(\gamma^{-1}\gamma^\prime)$
for every $f\in L^2(\widehat{\mathcal{U}_A})$ ($\gamma,\gamma^\prime\in\widehat{\mathcal{U}_A}$)
\footnote{$E_{U}(V)$ can be rewritten as the Fourier transform of the Kac-Takesaki operator $W_U$
defined by $(W_U v)(u):=U_uv(u)$ for every $v\in \mathcal{H}_\omega \otimes L^2(\mathcal{U}_A)$
($u\in \mathcal{U}_A$) : $E_{U}(V)=(id\otimes\mathcal{F)}W_U^{\ast}(id\otimes\mathcal{F}^{-1})$.
This rewrite is essential for us to understand the Fourier analytic aspect of measurement processes,
which is emphasized in \cite{HO09}.}.
A measuring interaction $\tau_U(V)$ of the ideal measurement of $A$ is then defined by
\begin{equation*}
\tau_U(V)(X)=E_{U}(V)^\ast XE_{U}(V),\hspace{5mm}
X\in \pi_\omega(\mathcal{X})^{\prime\prime}\otimes L^\infty(\widehat{\mathcal{U}_A}).
\end{equation*}

\section{Generalized Born Rule}
In this section we consider the case $M=\widehat{\mathcal{U}_A}$.

The spectral distribution of an observable is crucial information of the state of the system  
under an experimental situation, the latter of which is in nature macroscopic. 
Hence it is quite natural and important to consider 
how and in which form we can deduce the Born rule, 
the most fundamental rule for spectral distribution in quantum physics, 
from mathematical framework even applicable to such a context with infinite degrees of freedom. 
In the present section we formulate ``generalized Born rule'' 
on the basis of measurement processes in terms of sectors. 


Let $\mathfrak{m}=(\alpha_{\mathfrak{m}},\psi_\mathfrak{m})$ be a measurement process, 
where $\psi_\mathfrak{m}$ is a normal state. Consider the GNS representation 
$(\pi_{\varphi,\mathfrak{m}},\mathcal{H}_{\varphi,\mathfrak{m}},\Omega_{\varphi,\mathfrak{m}})$ of
 $(\mathcal{A}\otimes C_0(\widehat{\mathcal{U}_A}), \varphi_\mathfrak{m})$ given by 

\begin{align*}
\mathcal{H}_{\varphi,\mathfrak{m}} &=
\overline{(\pi_\omega\otimes id)(\mathcal{A}\otimes C_0(\widehat{\mathcal{U}_A}))U_{\mathfrak{m}}
(\xi_\varphi\otimes\xi_{\psi_\mathfrak{m}})},\\
 \pi_{\varphi,\mathfrak{m}}(X)&=(\pi_\omega\otimes id)(X) |_{\mathcal{H}_{\varphi,\mathfrak{m}}},\hspace{5mm}
 X\in \mathcal{A}\otimes C_0(\widehat{\mathcal{U}_A}),\\
 \Omega_{\varphi,\mathfrak{m}}&=U_{\mathfrak{m}}(\xi_\varphi\otimes\xi_{\psi_\mathfrak{m}}).
\end{align*}
where $U_{\mathfrak{m}}$ and $\xi_\varphi$, $\xi_{\psi_\mathfrak{m}}$ are a unitary implementer of $\alpha_{\mathfrak{m}}$ and $\xi_\varphi$, $\xi_{\psi_\mathfrak{m}}$ of vector representations of $\varphi$, $\psi_\mathfrak{m}$, respectively, 
defined in the standard representation \cite{BR1} of a von Neumann algebra $\pi_\omega(\mathcal{A})^{\prime\prime}\otimes L^\infty(\widehat{\mathcal{U}_A})$.

The von Neumann algebra corresponding to this representation is
$\pi_{\varphi,\mathfrak{m}}(\mathcal{A}\otimes C_0(\widehat{\mathcal{U}_A}))^{\prime\prime}$, for which
\[\pi_{\varphi,\mathfrak{m}}(\mathbb{C}1\otimes C_0(\widehat{\mathcal{U}_A}))^{\prime\prime}\subset
\mathcal{Z}_{\pi_{\varphi,\mathfrak{m}}}(\mathcal{A}\otimes C_0(\widehat{\mathcal{U}_A}))\]
holds. Moreover,
\[
(1\otimes\chi_\Delta )|_{\mathcal{H}_{\varphi,\mathfrak{m}}} \in \pi_{\varphi,\mathfrak{m}}(\mathbb{C}1\otimes C_0(\widehat{\mathcal{U}_A}))^{\prime\prime}
\]
holds for any Borel set $\Delta$ of states on $\mathcal{A}\otimes C_0(\widehat{\mathcal{U}_A})$.

Let us consider the subcentral decomposition 
\[
\varphi_\mathfrak{m}(X)=\int d^\mathcal{B}\varphi_\mathfrak{m}(\rho)\;\rho(X)
\]
for any von Neumann algebra $\mathcal{B}$ satisfying
\[
\pi_{\varphi,\mathfrak{m}}(\mathbb{C}1\otimes C_0(\widehat{\mathcal{U}_A}))^{\prime\prime}
\subset \mathcal{B}\subset \mathcal{Z}_{\pi_{\varphi,\mathfrak{m}}}(\mathcal{A}\otimes C_0(\widehat{\mathcal{U}_A})).
\]

Then the subcentral decomposition of 
$\varphi_\mathfrak{m}$ for $\mathcal{B}$ becomes an extremal decomposition
(i.e. decomposition into pure states) of $\varphi_\mathfrak{m}|_{\mathbb{C}1\otimes C_0(\widehat{\mathcal{U}_A})}$,
where $\cdot|_{\mathbb{C}1\otimes C_0(\widehat{\mathcal{U}_A})}$ 
means the restriction of states on $\mathcal{A}\otimes C_0(\widehat{\mathcal{U}_A})$
 into $\mathbb{C}1\otimes C_0(\widehat{\mathcal{U}_A})$. 
Hence each $\rho\in\text{supp}\;d^\mathcal{B}\varphi_\mathfrak{m}$ can be represented as $\varphi^\prime\otimes\delta_\gamma$ $(\varphi^\prime \in E_{\mathcal{A}}, \gamma\in \widehat{\mathcal{U}_A})$.

Then the instrument functional for the subcentral measure $d^\mathcal{B}\varphi_\mathfrak{m}$ corresponding to $\mathcal{B}$ is defined as 
\[
\mathcal{I}(f;d^\mathcal{B}\varphi_\mathfrak{m})(X)=\int d^\mathcal{B}\varphi_\mathfrak{m}(\rho)\;f(\rho)\;\rho(X)
\]
for any $X\in \mathcal{A}\otimes C_0(\widehat{\mathcal{U}_A})$ and $f\in L^\infty(d^\mathcal{B}\varphi_\mathfrak{m})$.

Especially, 
for the function $\check{\chi}_\Delta:=\chi_{\{\rho \in E_{\mathcal{A}\otimes C_0(\widehat{\mathcal{U}_A})}| \rho = \varphi ' \otimes \delta_{\gamma},  \varphi '\in E_{A},  \gamma \in \Delta \}}$ 
corresponding to a Borel set $\Delta\subset \widehat{\mathcal{U}_A}$\footnote{$\check{\chi}_\Delta$ can be considered as an indicator function on the state space 
by applying Gelfand-Naimark theorem for non-unital C*-algebras and locally compact spaces. }, 
\begin{align*}
\mathcal{I}(\check{\chi}_\Delta; d^\mathcal{B}\varphi_\mathfrak{m})(\tilde{X}\otimes 1)
&=\int d^\mathcal{B}\varphi_\mathfrak{m}(\rho)\;\check{\chi}_\Delta(\rho)\;\rho(\tilde{X}\otimes 1) \nonumber\\
 &= (\tilde{\varphi}\otimes \psi_\mathfrak{m})(\alpha_\mathfrak{m}(\pi_\omega(\tilde{X})\otimes \chi_\Delta)),
\end{align*}
since the inner product is available in the standard representation. For the case $\tilde{X}=1$, we obtain the following: 

\begin{theo}[Generalized Born rule]\label{GBorn}
Let $\varphi=\tilde{\varphi}\circ\pi_\omega$ be a normal state for object system and $\mathfrak{m}=(\alpha_{\mathfrak{m}},\psi_\mathfrak{m})$ be 
a measurement of an observable $A\in \pi_\omega(\mathcal{A})^{\prime\prime}$. The probability that the output belongs to a Borel set $\Delta$ of $\widehat{\mathcal{U}_A}$ (or $\mathbb{R}$) is given by
\[
(\tilde{\varphi}\otimes \psi_\mathfrak{m})(\alpha_\mathfrak{m}(1\otimes \chi_\Delta)).
\]
\end{theo}
\vspace{3mm}

\begin{rema}
Our argument can be generalized for sigular states such as ``neutral position'' (a delta measure on $L^\infty(\widehat{\mathcal{U}_A})$), which is an ideal state of measurement apparatus. 
Let 
$\{\mathfrak{m}_j=(\alpha_\mathfrak{m},\psi_{\mathfrak{m}_j})$ be a net such that $\psi_{\mathfrak{m}_j}\rightarrow \psi_\mathfrak{m}$ (in $\sigma$-weak topology). Then 
\[
\lim_{j} \mathcal{I}(\check{\chi}_\Delta;d^\mathcal{B}\varphi_{\mathfrak{m}_j})(X\otimes 1)
=(\tilde{\varphi}\otimes\psi_\mathfrak{m})[\alpha_\mathfrak{m}(\pi_\omega(X)\otimes \chi_\Delta)]
\]
Hence the generalized Born rule itself holds even for singular states, although $\sigma$-additivity of $(\tilde{\varphi}\otimes \psi_\mathfrak{m})(\alpha_\mathfrak{m}(1\otimes \chi_\Delta))$ may fail to hold in general. 

\end{rema}

\section{Born Rule and Spectral Equivalence}

In the present section we see how and under which condition the usual Born rule can be derived.

Let us first explain the mathematical concept of spectral equivalence, 
which is formulated by Ozawa \cite{Oz05,Oz06} under the name of ``perfect correration'' and generalized by KO in \cite{OjOk13}. 
Consider a ($\sigma$-finite) von Neumann algebra $\mathcal{M}$ on a Hilbert space $\mathcal{H}$.

Two projection-valued measures $E_1,E_2$ on a mesurable space $(S,\mathcal{B}(S))$ are said to be spectrally equivalent on a family $\mathcal{S}$ of 
states and 
denoted as $E_1=_\mathcal{S}E_2$ if
\begin{equation}
\varphi(E_1(\Delta)E_2(\Gamma))=0
\end{equation}
holds for any $\varphi\in\mathcal{S}$ and any pair of Borel sets $\Delta,\Gamma$ satisfying $\Delta\cap\Gamma=\emptyset$.
We also use a simpler notation $E_1=_\varphi E_2$ when $\mathcal{S}=\{\varphi\}$.

Self-adjoint elements $X, Y\in \mathcal{M}$ are also 
said to be spectrally equivalent on a family $\mathcal{S}$ and denoted as $X=_\mathcal{S}Y$ if so are the corresponding projection-valued measures.

Let $\mathcal{S}$ be a subset of normal states. In the standard representation there exists $\xi_\varphi$ in the corresponding to a normal state $\varphi$ such that
\[
\varphi(A)=\left\langle \xi_\varphi, X\xi_\varphi \right\rangle,\hspace{5mm}X\in\mathcal{M}.
\]
In this case it is easy to see that spectral equivalence is indeed an equivalence relation;
\begin{center}
(1) $X=_\mathcal{S}X$;\hspace{4mm}(2) $X=_\mathcal{S}Y\Rightarrow Y=_\mathcal{S}X$;
\hspace{4mm}(3) $X=_\mathcal{S}Y,\;Y=_\mathcal{S}Z\Rightarrow X=_\mathcal{S}Z$.
\end{center}
(1) and (2) are obvious; (3) can be proved by appealing the fact that for 
any $\varphi\in\mathcal{S}$ there exists $\xi_\varphi$ 
satisfying $\varphi(A)=\left\langle \xi_\varphi, X\xi_\varphi \right\rangle$ 
and $(E^X(\Delta)-E^Y(\Delta))\xi_\varphi=0$ for any Borel set $\Delta$.

Moreover the theorem below holds, which is a generalization of Theorem 5.3 in \cite{Oz06}.
\begin{theo}\label{PC2}
$X=_\varphi Y$ if and only if for any pair of Borel sets $\Delta,\Gamma$ 
the joint probability measure on $\mathbb{R}^2$ is supported by a closed subset of the 
diagonal set $\mathbb{D}=\{(x,x)\in\mathbb{R}^2|x\in\mathbb{R}\}$. 
\end{theo}

The theorem above is useful to clarify the meaning of spectral equivalence in the context of measurement theory:
For measurement (classical or quantum) it is desirable that the spectrum of $A\otimes 1$ 
before the measurement can be considered as a subset of $\Delta$ when the spectrum 
detected by $\alpha_\mathfrak{m}(1\otimes \chi_{(\cdot)})$ is equal to $\Delta\in\mathcal{B}(\mathbb{R})$.
Moreover, the spectrum of $A\otimes 1$ should be stable under the change of representation in a sector. 
These assumptions are summarized as the equation below:
\begin{equation*}\label{MPPC}
\alpha_\mathfrak{m}(1\otimes \chi_{(\cdot)})=_{\pi_\omega(\mathcal{A})^{\prime\prime}_{\ast,1}\otimes
\psi_\mathfrak{m}} E^A(\cdot)\otimes 1.
\end{equation*}
where
$\pi_\omega(\mathcal{A})^{\prime\prime}_{\ast,1}\otimes \psi_\mathfrak{m}=\{\varphi\otimes \psi_\mathfrak{m}|
\varphi \text{ is a normal state on }
\pi_\omega(\mathcal{A})^{\prime\prime}\}$. 
When the equation above holds we obtain the following fundamental theorem from Theorem \ref{PC2}.

\setcounter{section}{1}
\begin{theor}[Born rule]\label{Born}
Assume that the state of the object system is $\varphi=\tilde{\varphi}\circ\pi_\omega$.                                                                                                                                                                                                                                                                         
Then the probabilty $\mathrm{Pr}\{A\in\Delta\Vert\varphi\}$ with which 
the value of $A\in\pi_\omega(\mathcal{X})^{\prime\prime}$ is 
in $\Delta\in\mathcal{B}(\mathbb{R})$ is given by 
\begin{equation*}
\tilde{\varphi}(E^A(\Delta))=\langle \xi_{\tilde{\varphi}},E^A(\Delta)\xi_{\tilde{\varphi}} \rangle .
\end{equation*}
\end{theor}

\end{document}